\documentclass{template}

\usepackage{orcidlink}
\usepackage{amsmath}
\usepackage{amssymb}
\usepackage{textcomp}
\hypersetup{hidelinks}
\providecommand{\ket}[1]{\lvert #1 \rangle}
\providecommand{\bra}[1]{\langle #1 \rvert}
\providecommand{\braket}[2]{\langle #1 \vert #2 \rangle}

\newcommand{\UFSCar}{Departamento de F\'{i}sica, Universidade Federal de S\~{a}o Carlos, Rod. Washington Lu\'{i}s, km 235 - SP-310, 13565-905 S\~{a}o Carlos, SP, Brazil}

\newcommand{\Unesp}{Universidade Estadual Paulista (UNESP), Instituto de Ci\^{e}ncias e Engenharia, 18409-010 Itapeva, S\~{a}o Paulo, Brazil}

\newcommand{\Kipu}{Kipu Quantum GmbH, Greifswalderstrasse 212, 10405 Berlin, Germany}

\begin{document}

% header
\pagestyle{fancy}
\fancyhf{}
\fancyhead[R]{\small
    \href{https://doi.org/10.1002/andp.70283}
    {https://doi.org/10.1002/andp.70283}}

\renewcommand{\headrulewidth}{0.4pt}
\renewcommand{\footrulewidth}{0pt}

\fancyfoot[C]{\thepage}

\title{Quantum Limits to Linewidth Narrowing in Single- and Few-Atom Cavity Electromagnetically Induced Transparency}

\maketitle
\normalsize

\begin{center}

{\normalsize
Lucas R. S. Santos~\orcidlink{0000-0001-9700-884X}\textsuperscript{1,*},
Murilo H. Oliveira~\orcidlink{0000-0001-9777-8342}\textsuperscript{2},
Luiz O. R. Solak~\orcidlink{0000-0002-4760-3357}\textsuperscript{1},
Daniel Z. Rossatto~\orcidlink{0000-0001-9432-1603}\textsuperscript{3},
and Celso J. Villas-Boas~\orcidlink{0000-0001-5622-786X}\textsuperscript{1}
}

\vspace{0.8em}

{\footnotesize
\textsuperscript{1}\UFSCar\\
\textsuperscript{2}\Kipu\\
\textsuperscript{3}\Unesp\\
\textsuperscript{*}Corresponding author:
\href{mailto:lucasrss@estudante.ufscar.br}
{lucasrss@estudante.ufscar.br}
}

\end{center}

\normalsize


\begin{abstract}
Electromagnetically induced transparency (EIT) in cavities can narrow the transmission resonance below the empty-cavity linewidth. We investigate the limits of this narrowing from a single emitter to the few-atom regime. Using a Lindblad master equation for $N_\text{at}$ identical three-level atoms coupled to a cavity mode, driven by coherent probe and control fields, we compute the full width at half maximum (FWHM) of the transparency feature. In the strictly low-excitation limit, we derive an analytical cubic polynomial that captures the narrowing and recovers the known linear-response scaling. For finite probe powers, the achievable linewidth faces a quantum bound on the minimum achievable linewidth. Contrasting our model with a semiclassical approximation, we show this limitation arises from the unavoidable excitation of higher-order multiphoton states. Their intrinsically larger decay rates destroy the ideal single-excitation EIT dark state. Increasing $N_\text{at}$ enhances collective cooperativity, creating a multiphoton blockade that suppresses these detrimental excitations and yields a stepwise reduction of the minimum FWHM. Our results provide analytical boundaries for cavity-EIT linewidth control, guiding the optimization of narrowband filters and highly coherent light–matter interfaces.
\end{abstract}

\section{Introduction}

Electromagnetically induced transparency (EIT) \cite{PhysRevLett.66.2593, RevModPhys.77.633} is a quantum-interference effect that enables coherent control of absorption and dispersion in multilevel systems. Beyond its role in slow-light experiments \cite{hau1999light, zhang2007slow}, EIT has become a versatile resource for quantum memories and light--matter interfaces, including protocols targeting non-classical states of light \cite{PhysRevLett.86.783, liu2001observation, specht2011single, LEI_Review_EIT}, as well as for EIT-based laser cooling of trapped particles \cite{PhysRevLett.85.4458, PhysRevLett.85.5547, PhysRevA.93.053401}. Moreover, by enabling large optical nonlinearities at low light levels, EIT underpins few-photon nonlinear optics and optical switching proposals and demonstrations \cite{PhysRevLett.64.1107, lukin2001controlling, werner1999photon, mucke2010electromagnetically}, and—in cavity and circuit implementations—offers a platform to engineer narrowband resonator responses and quantum fluctuations relevant to quantum networking, metrology, and device optimization \cite{Lukin:98, Wang:00, PhysRevLett.111.113602, RevModPhys.Rempe, PhysRevLett.104.193601}. Recent advances in deterministic trapping and positioning of few-atom arrays in optical cavities further motivate revisiting cavity-EIT controllability beyond the single-emitter limit \cite{PRXQuantum.6.010322}. Consequently, EIT is now widely regarded as an enabling mechanism for second-generation quantum technologies.

A particularly important consequence of EIT is the possibility of engineering very narrow optical resonances while maintaining strongly suppressed absorption. In a cavity, the steep EIT dispersion modifies the effective round-trip phase and group delay of the intracavity field, reshaping the resonator transmission and allowing the central transparency resonance to become narrower than the empty-cavity linewidth. Such linewidth narrowing is not only a spectroscopic feature. By producing a field-controllable narrow resonance, it improves spectral selectivity and enables tunable cavity responses for high-resolution spectroscopy and frequency stabilization \cite{Lukin:98, Wang:00}. The associated enhancement of the intracavity group delay and effective photon dwell time is also useful for cavity-ringdown and cavity-enhanced sensing schemes \cite{Yang:04}. In addition, the reduced spectral bandwidth provides a natural resource for narrowband light-matter interfaces, including cavity-enhanced quantum memories \cite{ma2022high} and few-emitter cavity-QED platforms relevant to quantum networking and controlled photon statistics \cite{mucke2010electromagnetically,Lei2023}.

In EIT experiments, a weak probe field can be transmitted through an otherwise opaque optical medium~\cite{PhysRevLett.105.153604}, or even a single atom \cite{mucke2010electromagnetically}, when a strong control field is applied to it \cite{RevModPhys.77.633}. Specifically, EIT arises under appropriate conditions involving the Rabi frequencies of the probe ($\Omega_p$) and control ($\Omega_c$) fields, leading to the coherent population trapping (CPT) phenomenon. This is characterized by the emergence of an atomic dark state, defined as a state decoupled from the light fields, consisting of a coherent superposition of the two ground states of the atomic system in the $\Lambda$-configuration \cite{agap1993coherent}, for example. Under the condition $\Omega_p\ll \Omega_c$, the absorption of a weak probe field tuned resonantly with some atomic transition is suppressed \cite{RevModPhys.77.633}, giving rise to a significant optical nonlinearity in the susceptibility and a corresponding change in the refractive index \cite{PhysRevLett.64.1107, lukin2001controlling}. 

The progress of linewidth narrowing in EIT has developed from semiclassical intracavity proposals to few-emitter cavity-QED implementations. The possibility of narrowing a cavity resonance by exploiting intracavity EIT was theoretically introduced in Ref.~\cite{Lukin:98}, where the strong dispersion of the transparent medium was shown to produce frequency pulling and linewidth narrowing. This effect was then experimentally observed in a ring cavity containing an EIT medium, where the linewidth was controlled through the coupling-field power \cite{Wang:00}. Subsequent works explored related mechanisms for enhancing cavity ringdown response \cite{Yang:04}, achieving narrow tunable resonances through interacting dark states in tripod configurations \cite{Ying:14}, and interpreting the narrowed cavity response in terms of dark-state polaritons \cite{Lin_2016}. More recently, cavity-EIT concepts have continued to support cavity-enhanced quantum memories and driven many-body cavity-QED platforms \cite{ma2022high, Lei2023}.

A driven cavity coupled to the atomic medium can replace the probe field applied directly to it, giving rise to cavity EIT \cite{mucke2010electromagnetically,werner1999photon, zhang2007slow,PhysRevLett.104.193601,PhysRevLett.105.153603}. In this extension, the nonlinearity introduced by EIT can be employed in other kinds of application, for instance to control the linewidth of optical cavities, enabling a linewidth smaller than the natural one of the cavity \cite{mucke2010electromagnetically,ficek1997spectral,Lukin:98, borges2017influence, PhysRevA.104.063704,Wang:00}. In the single-atom and few-atom cavity-QED regime, Ref.~\cite{mucke2010electromagnetically} showed that the linewidth of the central transparency resonance, quantified by the full width at half maximum (FWHM) of the cavity transmission spectrum, can be tuned by the control field and scales as $\Omega_c^2$ in the EIT regime. Therefore, the cavity linewidth can be reduced by decreasing the Rabi frequency of the control field. However, this reduction cannot be made arbitrarily, because excessively decreasing $\Omega_c$ causes the system to leave the EIT regime, and consequently the aforementioned expression derived to describe the FWHM scaling is no longer valid.

In this work, we theoretically investigate the control of linewidth narrowing within the cavity-EIT framework by varying $\Omega_c$ across all regimes. We show that, although the control field is crucial for adjusting the cavity response, there exists an optimal control field strength that yields a minimum achievable linewidth. Deviating from this optimum, either towards weaker or stronger control fields, causes the linewidth to broaden, recovering the empty-cavity response for $\Omega_c \to 0$ or $\Omega_c \to \infty$. We also analyze how the probe-field amplitude, the atom--field coupling strength, and the number of trapped atoms influence the narrowing of the cavity linewidth.

While the influence of decay and light-mediated mechanisms on EIT-based transparency features has been investigated both theoretically \cite{Lukin:98, borges2017influence, PhysRevA.104.063704,Lin_2016} and experimentally \cite{Wang:00,mucke2010electromagnetically,Ying:14}, the ultimate physical limitations to cavity linewidth narrowing in the few-emitter quantum regime remain less explicit. Here we address this question using a quantum master equation description of cavity EIT, systematically mapping the cavity transmission linewidth as a function of the control field strength $\Omega_c$, probe amplitude $\varepsilon$, atom-field coupling $g$, and the number of trapped atoms $N_\text{at}$. In the low-excitation regime, we derive an exact analytical cubic polynomial that perfectly captures the EIT linewidth and recovers the known linear-response scaling limits. Away from this ideal limit, we find that the achievable linewidth narrowing exhibits an optimal operating point and is ultimately bounded by a quantum limit. By contrasting our quantum model with a semiclassical approximation, we demonstrate that this limitation arises directly from the unavoidable excitation of higher-order multiphoton states. These discrete multiphoton components possess intrinsically larger decay rates that destroy the ideal single-excitation EIT dark state, thereby establishing how genuinely quantum nonlinear effects bound the ultimate controllability in cavity-EIT systems.

The structure of this paper is organized as follows. Section~\ref{sec:PHYSICAL SYSTEM AND MODEL} introduces the theoretical model for
$N_{\text{at}}$ identical $\Lambda$-type atoms coupled to a single cavity mode. In Section~\ref{sec:TRANSMISSION IN CAVITY-EIT}, we focus on the
single-atom case ($N_{\text{at}}=1$), deriving the exact analytical linewidth polynomial for the low-excitation regime and analyzing the breakdown of the narrowing control due to higher probe intensities and varying atom--field coupling. We then examine the few-atom regime ($N_{\text{at}}>1$) in Section~\ref{sec:few_atom_regime}, highlighting how increasing $N_{\text{at}}$ enhances the collective nonlinearity, leading to a strong multiphoton blockade. This progressive suppression of the detrimental multiphoton excitations yields a stepwise linewidth narrowing, explicitly unveiling the physical origin of the quantum limit. Finally, Section~\ref{sec:CONCLUSIONS} summarizes our findings and discusses their implications.

\section{Physical system and model}\label{sec:PHYSICAL SYSTEM AND MODEL}

\begin{figure}[t!]
    \centering
    \includegraphics[width=1.0\linewidth]{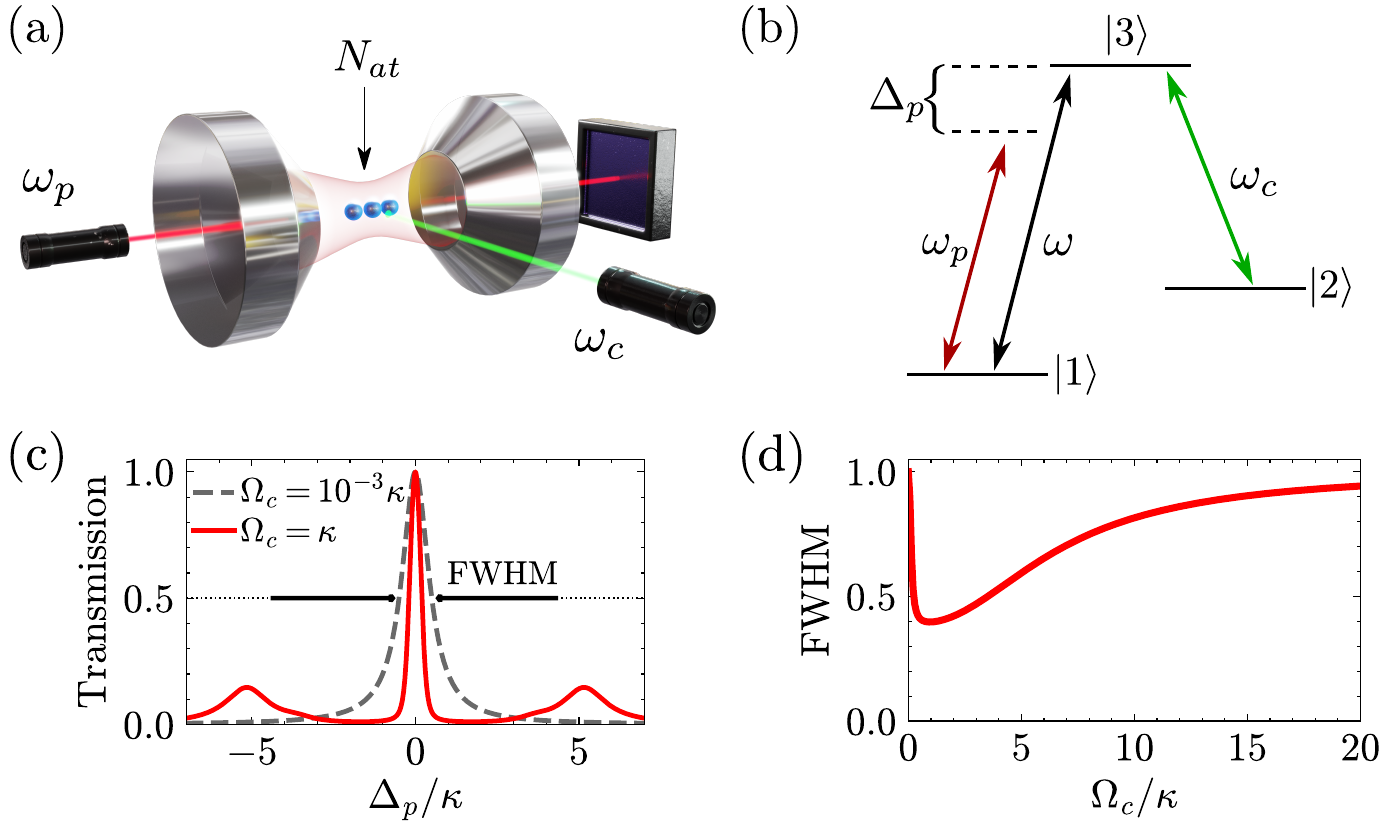}
    \caption{(a) Cavity-EIT setup with $N_{\text{at}}$ atoms coupled to the cavity mode. A probe field drives the cavity, while a control field drives the atoms. (b) $\Lambda$-type three-level atom with two ground states ($\ket{1}$ and $\ket{2}$) and an excited one ($\ket{3}$). The transition $\ket{2} \leftrightarrow \ket{3}$ is driven resonantly by the control field with frequency $\omega_c$, while the transition $\ket{1} \leftrightarrow \ket{3}$ is resonantly coupled to the cavity mode with frequency $\omega$. Moreover, a probe field, with frequency $\omega_p$, pumps the cavity mode with a detuning $\Delta_p = \omega_p - \omega$. (c) Normalized transmission spectrum of the cavity as a function of the normalized probe detuning $\Delta_p/\kappa$. Throughout this work, the full width at half maximum (FWHM) is expressed in units of the cavity field decay rate $\kappa$. For almost zero control field strength ($\Omega_c=10^{-3}\kappa$), the transmission spectrum becomes equivalent to an empty-cavity scenario $(\text{FWHM} = 1.0)$, on the other hand we have the cavity-EIT spectrum for $\Omega_c = \kappa$. (d) FWHM dependence with the normalized Rabi frequency of the control field $\Omega_c/\kappa$. In the last two panels we used a single atom ($N_{\text{at}}=1$), probe field strength $\varepsilon=\sqrt{0.1}\kappa$, atom-field coupling strength $g=5\kappa$ and atomic spontaneous decay rates  $\Gamma_{31}=\Gamma_{32}=0.5\kappa$.} 
\label{fig:scheme}
\end{figure}

Figure~\ref{fig:scheme} illustrates the physical system under consideration and its characteristic cavity-EIT transmission properties. The general setup we investigate is depicted in Figure~\ref{fig:scheme}a, where one or more identical three-level noninteracting atoms are confined within a two-sided optical cavity pumped by a coherent probe laser. As shown in Figure~\ref{fig:scheme}b, the atomic energy level has a $\Lambda$-type configuration. Throughout this work, we consider the resonant cavity-EIT configuration. The cavity mode with frequency $\omega$ is taken resonant with the $\ket{1}\leftrightarrow\ket{3}$ transition, so that the atom-cavity detuning is set to zero. The classical control field, with frequency $\omega_c$, is also taken resonant with the $\ket{2}\leftrightarrow\ket{3}$ transition. The control field induces a Rabi frequency $2\Omega_c$, while the coherent probe field, with frequency $\omega_p$, pumps the cavity mode with driving strength $\varepsilon$. Thus, the only detuning kept explicitly in our model is the probe-cavity detuning $\Delta_p=\omega_p-\omega$, which is scanned to obtain the cavity transmission spectrum and the corresponding FWHM. With this convention, $\Delta_p=0$ corresponds to the two-photon resonance condition and to the center of the cavity-EIT transparency window. This resonant choice allows us to isolate the linewidth-narrowing mechanism from additional frequency shifts or asymmetries induced by finite atom-cavity or control-field detunings. Within the electric dipole and rotating-wave approximations, the time-independent Hamiltonian, written in a frame rotating with the probe frequency $\omega_p$, is
($\hbar=1$)~\cite{PhysRevLett.111.113602}
\begin{equation}
     H = \Delta_p S_{11} - \Delta_p a^\dagger a + (\varepsilon a + g a S_{31} + \Omega_c S_{32} + \text{H.c.}),
     \label{Hamiltonian}
\end{equation}
where $\Delta_p = \omega_p - \omega$ is the probe-cavity detuning and
$S_{ij}=\sum_{k=1}^{N_{\text{at}}}\sigma_{ij}^{(k)}$ are collective atomic operators
with $\sigma_{ij}^{(k)}=\ket{i}^{(k)}\!\bra{j}^{(k)}$. For $i\neq j$, $\sigma_{ij}^{(k)}$ are the raising and lowering operators of the $k$th atom, while $\sigma_{ii}^{(k)}$ are population operators. Respectively, $a$ and $a^\dagger$ are the photon annihilation and creation operators, while $\text{H.c.}$ stands for Hermitian conjugate.

In all of our simulations, we assume a uniform atom--field coupling $g$ for each atom, corresponding to atoms trapped at well-defined (and equivalent) positions within the cavity mode. This assumption is consistent with experimental platforms that achieve robust confinement and deterministic
positioning of few-atom arrays in optical cavities, as demonstrated in Ref.~\cite{PRXQuantum.6.010322}. The system dynamics, including the dissipative effects at $T = 0 \, \text{K}$, is determined by the master equation \cite{PhysRevLett.111.113602} 
\begin{equation}
\dfrac{d \rho}{d t} = -i[H, \rho] + \dfrac{\kappa}{2}(2a\rho a^\dagger - a^\dagger a\rho - \rho a^\dagger a)
 + \sum\limits_{k=1}^{N_{\text{at}}} \sum\limits_{l=1}^2 \left[\dfrac{\Gamma_{3l}}{2}\left( 2\sigma_{l3}^{(k)}\rho\sigma_{3l}^{(k)} - \sigma_{33}^{(k)}\rho - \rho\sigma_{33}^{(k)} \right) \right],
\label{ME}
\end{equation}
where $\rho$ represents the density matrix of the atom--cavity system, $\kappa$ is the decay rate of the intracavity-field intensity, and $\Gamma_{3l}$ ($l=1,2$) are the spontaneous decay rates of the atomic transition $\ket{3}\rightarrow\ket{l}$.

The system dynamics is obtained numerically by solving the master equation in the steady state employing the Monte Carlo method \cite{Molmer:93}, with a number of trajectories large enough to reproduce the mean values derived through the master equation. In our simulations, we observed that more than 2000 trajectories are already enough to reproduce the master equation mean values. Thus, we have fixed 2560 trajectories in our simulations, being this number chosen to facilitate the use of parallel processing in the multicore cluster we have available. We truncate the Fock space dimension of the cavity mode to $N$ (which is chosen according to the probe field intensity) and use the Python library QuTip \cite{JOHANSSON20121760} to solve the dynamics of the system for a limited number of atoms ($N_{\text{at}}$), considering an initial state in which the cavity is in the vacuum state and each atom is in the second ground state, \textit{i.e.}, $\ket{\psi(0)}=\ket{0}_{\text{field}} \otimes \ket{2}_{\text{atom}}^{\otimes N_\text{at}}$. This limitation arises due to the exponential growth of the dimension of the density matrix $\rho$, $\text{dim}=M\times M$, where $M=3^{N_{\text{at}}}N$ \cite{nation2015steadystatesolutionmethodsopen}, thus limiting the numerical solution to a few atoms. By solving the master equation, one numerically calculates the normalized transmission spectrum of the cavity ($\langle a^\dagger a \rangle/|2\varepsilon/\kappa|^2$), atomic populations ($\langle S_{ii} \rangle$) and the intracavity photon number distribution ($ P_n = \langle \ket{n}\bra{n} \rangle$).

To characterize the strength of the atom-cavity interaction, we use the single-atom cooperativity parameter, defined as $C \equiv g^2/(2\kappa \Gamma)$, where $\Gamma = \Gamma_{31} + \Gamma_{32}$ is the total atomic decay rate. Although we are dealing with a three-level atom here, we adopt the same definition used for two-level atoms for the sake of simplicity and to facilitate comparison with other systems. The cooperativity parameter indicates the relative strength of coherent atom-photon coupling compared to dissipative processes. The inverse of $C$ gives the critical number of atoms required to significantly influence the cavity transmission. For two-level systems, we can also define the critical number of photons, $n_c=\Gamma^2/2g^2$, interpreted as the minimal number of photons needed to change the radiation properties of the atom \cite{RevModPhys.Rempe}. Although Equation~\ref{Hamiltonian} and \ref{ME} are written for an arbitrary number of atoms, we first focus on the single-atom case ($N_{\text{at}}=1$) in Section~\ref{sec:TRANSMISSION IN CAVITY-EIT},
where the mechanisms controlling linewidth are most clearly identified.
In Section~\ref{sec:few_atom_regime}, we then analyze how increasing $N_{\text{at}}$ modifies the photon statistics and the minimum achievable linewidth.

\begin{figure}[htb!]
    \centering
    \includegraphics[width=1.0\linewidth]{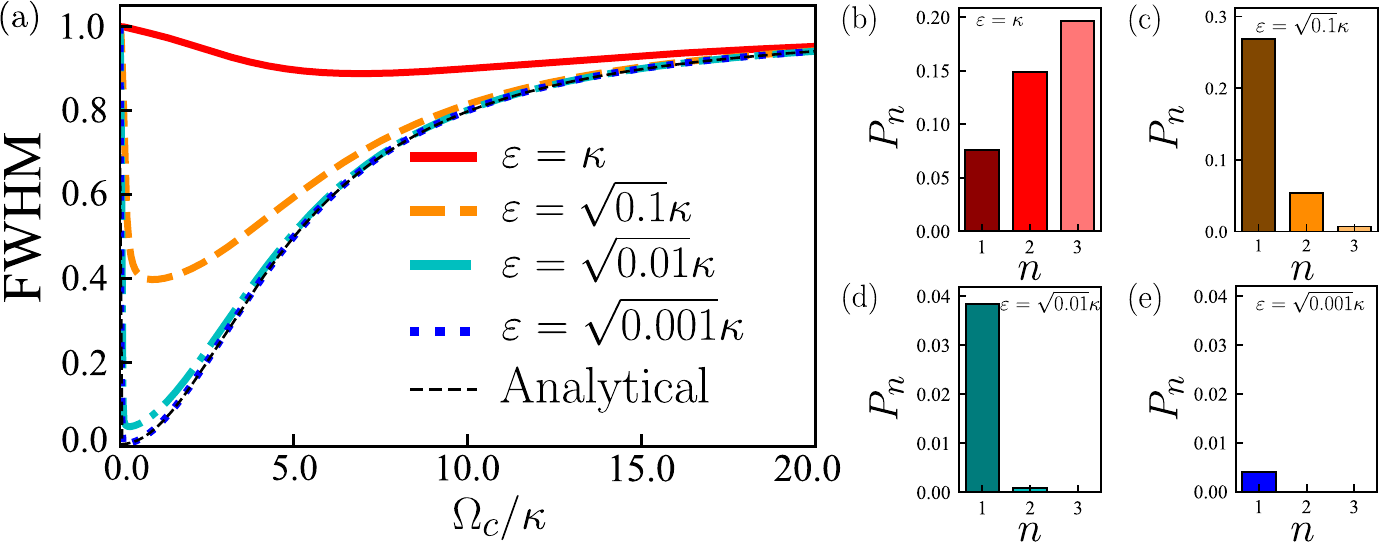}
    \caption{(a) FWHM of the central cavity-EIT resonance as a function of the normalized control-field strength $\Omega_c/\kappa$ for several probe strengths $\varepsilon$.  The dashed black curve shows the lowest real positive root of Equation~\ref{eq:fwhm_polynomial}. (b--e) Intracavity photon-number distribution $P_n$ at $\Delta_p=0$ and $\Omega_c=\kappa$ for (b) $\varepsilon=\kappa$, (c) $\varepsilon=\sqrt{0.1}\kappa$, (d) $\varepsilon=\sqrt{0.01}\kappa$, and (e) $\varepsilon=\sqrt{0.001}\kappa$. The distributions are shown up to $n=3$, since higher-photon components are negligible on the scale of the plots as the excitation is reduced from (b) to (e). In all panels, we set $N_{\text{at}}=1$, $g=5\kappa$, and $\Gamma_{31}=\Gamma_{32}=0.5\kappa$.}
    \label{fig:pn}
\end{figure}

\section{Transmission in cavity EIT}\label{sec:TRANSMISSION IN CAVITY-EIT}

\subsection{Dependence on the control field $\Omega_c$}\label{subsec:control_field_Oc}

The control field, with Rabi frequency $2\Omega_c$, is the main parameter for engineering the cavity transmission spectrum in the cavity-EIT configuration. Figure~\ref{fig:scheme}c illustrates two representative regimes for a single atom ($N_{\text{at}}=1$). For a very weak control field, $\Omega_c = 10^{-3}\kappa \ll g$, the atom remains effectively decoupled from the cavity mode and the transmission approaches the empty-cavity behavior. For $\Omega_c=\kappa$ (solid line), the characteristic cavity-EIT spectrum emerges: a narrow transparency resonance centered at $\Delta_p=0$, accompanied by two secondary peaks. These side peaks correspond to the Jaynes--Cummings \cite{JCM} dressed resonances shifted by the control field and are located around $\Delta_p\simeq \pm\sqrt{g^2 N_{\text{at}}+\Omega_c^2}$ \cite{PhysRevLett.111.113602}.

Throughout this work, we quantify the linewidth of the central transparency resonance by its full width at half maximum (FWHM), expressed in units of the cavity decay rate $\kappa$. The dependence of the FWHM on $\Omega_c$ for a representative probe strength is shown in Figure~\ref{fig:scheme}d. In the strict limit $\Omega_c\rightarrow 0$, the $\Lambda$ configuration is switched off and the system reduces to the bare-cavity response. For very large control fields, $\Omega_c\gg g$, the strong dressing also brings the spectrum back toward the empty-cavity linewidth. Between these limits, the system enters the cavity-EIT regime and there exists an operating point that yields the narrowest transparency window for the chosen parameters.

Figure~\ref{fig:pn} summarizes the behavior of the cavity-EIT linewidth and the corresponding intracavity photon-number distributions across different excitation regimes. A key aspect of linewidth control is that its effectiveness strongly depends on the excitation of the intracavity field. Figure~\ref{fig:pn}a compares the numerical FWHM as a function of $\Omega_c/\kappa$ for several probe-field strengths $\varepsilon$. In the strictly low-excitation regime, where multiphoton components are negligible, we can analytically determine the fundamental linewidth limits by treating the system in the single-excitation manifold. As derived in Appendix~\ref{app:analytical_fwhm}, combining a semiclassical approach with the ideal dark state population dynamics leads to an exact cubic polynomial for the FWHM, denoted by $F$:
\begin{equation}
    F^3 - \kappa F^2 - 4(\Omega_c^2 + A)F + 4\kappa\Omega_c^2 = 0,
    \label{eq:fwhm_polynomial}
\end{equation}
where $A = g^2 N_{\text{at}} \Omega_c^2 / (\Omega_c^2 + g^2 n_{\text{FWHM}})$ is the saturated collective coupling parameter and $n_{\text{FWHM}} = 2(\varepsilon/\kappa)^2$ is the intracavity photon number evaluated at the half-maximum condition of the empty cavity. The smallest real positive root of Equation~\eqref{eq:fwhm_polynomial} provides the exact semi-analytical linewidth, which is plotted as the dashed black line in Figure~\ref{fig:pn}a.

\begin{figure}[!htb]
\includegraphics[width = 1.0\textwidth, clip]{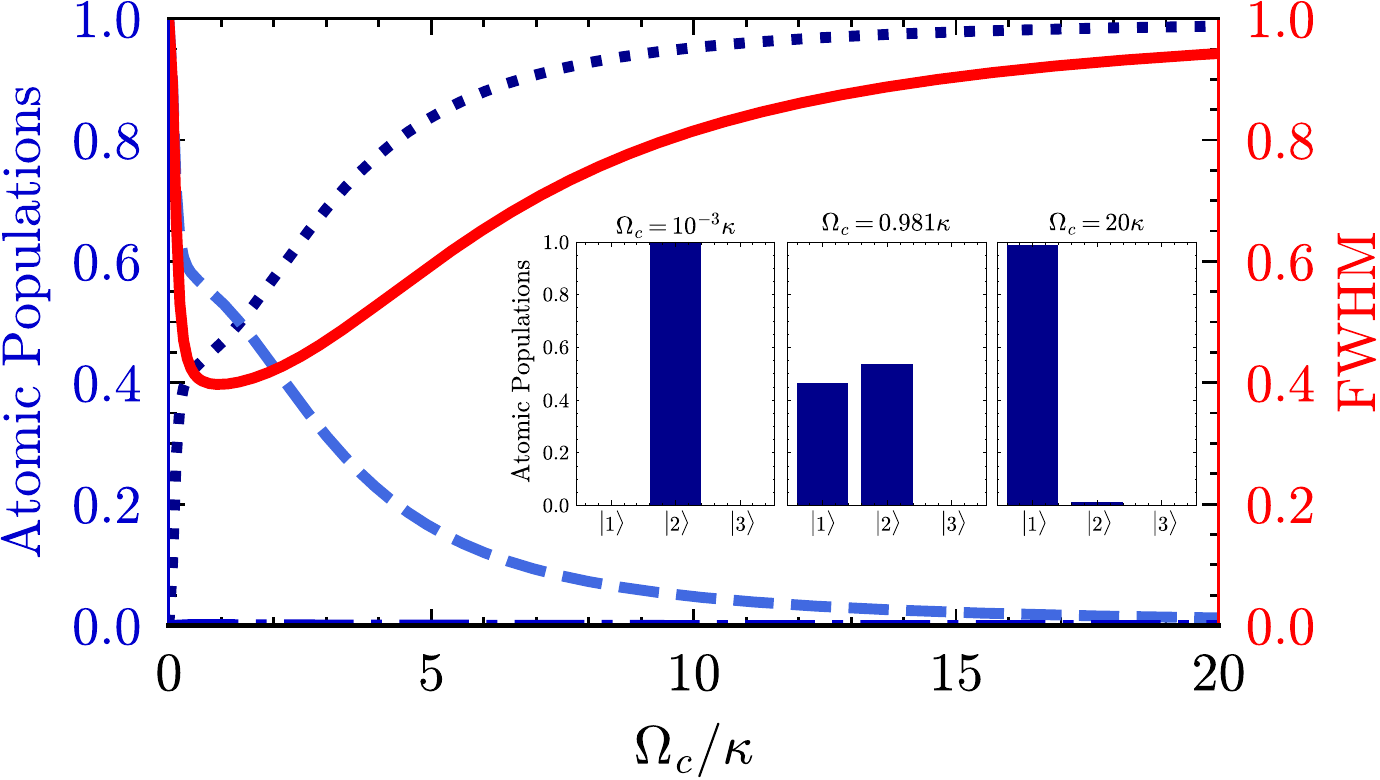}
    \caption{Atomic populations $\langle \sigma_{11}\rangle$, $\langle \sigma_{22}\rangle$, and $\langle \sigma_{33}\rangle$ (left axis, blue curves) and the corresponding FWHM of the central cavity-EIT resonance (right axis, red curve) as a function of the normalized control-field strength $\Omega_c/\kappa$ for $N_{\text{at}}=1$. The blue dotted, dashed, and dash-dotted curves correspond, respectively, to $\langle \sigma_{11}\rangle$, $\langle \sigma_{22}\rangle$, and $\langle \sigma_{33}\rangle$, while the red solid curve represents the FWHM. The excited-state population $\langle \sigma_{33}\rangle$ remains nearly zero over the whole range shown. For each value of $\Omega_c$, the atomic populations are evaluated at the edge of the corresponding transparency window, $\Delta_p=\text{FWHM}(\Omega_c)/2$. The inset shows the populations for three representative control strengths: $\Omega_c=10^{-3}\kappa$, $\Omega_c=\Omega_c^{\rm min}$, corresponding to the minimum FWHM, and $\Omega_c=20\kappa$. Here we set $g=5\kappa$, $\varepsilon=\sqrt{0.1}\kappa$, and $\Gamma_{31}=\Gamma_{32}=0.5\kappa$.}
    \label{fig:populations_extremes}
\end{figure}

Equation~\eqref{eq:fwhm_polynomial} provides a comprehensive framework to compare our findings with previous cavity-EIT linewidth narrowing studies. In vapor-cell and cold-atom implementations, as well as in single- and few-atom cavity-QED experiments, the linewidth narrowing is usually analyzed inside the ideal EIT regime. By taking the limit $\Omega_c \ll g\sqrt{N_{\text{at}}}$, our analytical polynomial naturally reduces to the familiar scaling $\mathrm{FWHM} \approx \kappa\Omega_c^2/(g^2N_\text{at})$ \cite{mucke2010electromagnetically, Lukin:98, Wang:00, Lin_2016}, directly recovering the $1/N_{\text{at}}$ narrowing. Moreover, our full solution connects this known behavior to the complementary strong-control limit ($\Omega_c\gg g\sqrt{N_\text{at}}$), where the control field dominates the dark state projection and the response approaches the empty-cavity linewidth, $F \to \kappa$.

 In contrast, for higher probe strengths, the achievable linewidth narrowing is strongly reduced and deviates from the analytical root, as seen from the $\varepsilon=\kappa$ and $\varepsilon=\sqrt{0.1}\kappa$ curves in Figure~\ref{fig:pn}a. This breakdown of linewidth control is a direct consequence of the inevitable appearance of multiphoton components in the intracavity field. As discussed in Appendix~\ref{app:analytical_fwhm}, higher-order multiphoton dressed states ($n \ge 2$) broaden the dark resonance. Therefore, the analytical solution given by Eq.~\eqref{eq:fwhm_polynomial} is physically valid only for very weak probes fields, $\varepsilon \lesssim \sqrt{0.001}\kappa$.

The photon-number distributions $P_n$ evaluated at the resonance ($\Delta_p=0$) and shown in Figure~\ref{fig:pn}b--\ref{fig:pn}e make this distinction physically explicit. In these panels, we plot $P_n$ only up to $n=3$, since for the lower-excitation curves the probability weight at $n>3$ is negligible. For $\varepsilon=\sqrt{0.01}\kappa$ and $\varepsilon=\sqrt{0.001}\kappa$ (Figure~\ref{fig:pn}d and e), the probability distribution is overwhelmingly concentrated in the single-photon component ($P_1$), confirming that the system operates effectively in the linear-optics regime where the analytical polynomial holds. Conversely, for $\varepsilon=\sqrt{0.1}\kappa$ and $\varepsilon=\kappa$ (Figure~\ref{fig:pn}c and b), the multiphoton components ($P_2$ and $P_3$) become significant. These terms destroy the ideal single-excitation EIT dark state, limiting the ability of a single atom to reshape the transmitted field and ultimately bounding the achievable minimum FWHM.

To connect the linewidth tuning to the atomic dynamics, Figure~\ref{fig:populations_extremes} plots the steady-state atomic populations together with the corresponding FWHM as functions of $\Omega_c/\kappa$, with the populations evaluated at the edge of the transparency window, $\Delta_p=\text{FWHM}(\Omega_c)/2$, for each $\Omega_c$. In the weak-control limit, the atom predominantly occupies $\ket{2}$ and effectively decouples from the cavity mode, while in the strong-control limit the population approaches $\ket{1}$. For the parameter set considered here, the minimum linewidth occurs at an intermediate control strength, where both ground states acquire appreciable population. The inset summarizes the populations at three representative values, $\Omega_c=10^{-3}\kappa$, $\Omega_c=\Omega_c^{\rm min}$, and $\Omega_c=20\kappa$. Importantly, the population balance at the minimum is not a universal signature of optimal narrowing; it depends on $g$, $\varepsilon$, and detunings, and it can shift away from an equal superposition when these parameters are varied.

\subsection{Dependence on the atom--field coupling $g$}\label{subsec:coupling_strength_g}

\begin{figure}[!htb]
\centering
\includegraphics[trim = 2.8mm 1.5mm 1.5mm 1mm, width = 0.49\textwidth, clip]{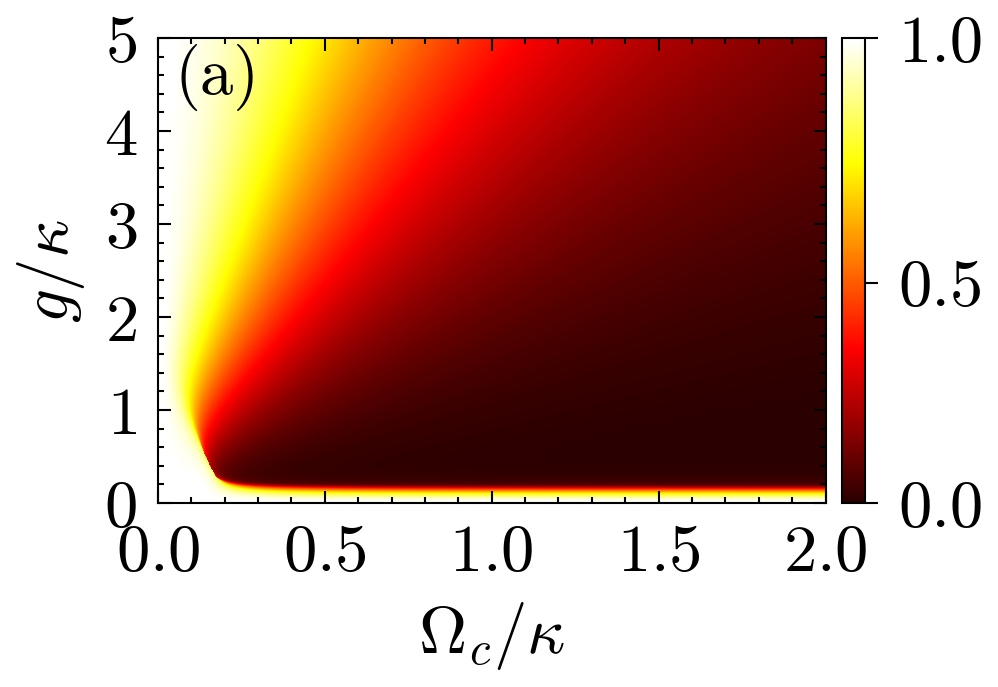} \quad
\includegraphics[trim = 1.0mm 1.5mm 2.2mm 1mm, width = 0.48\textwidth, clip]{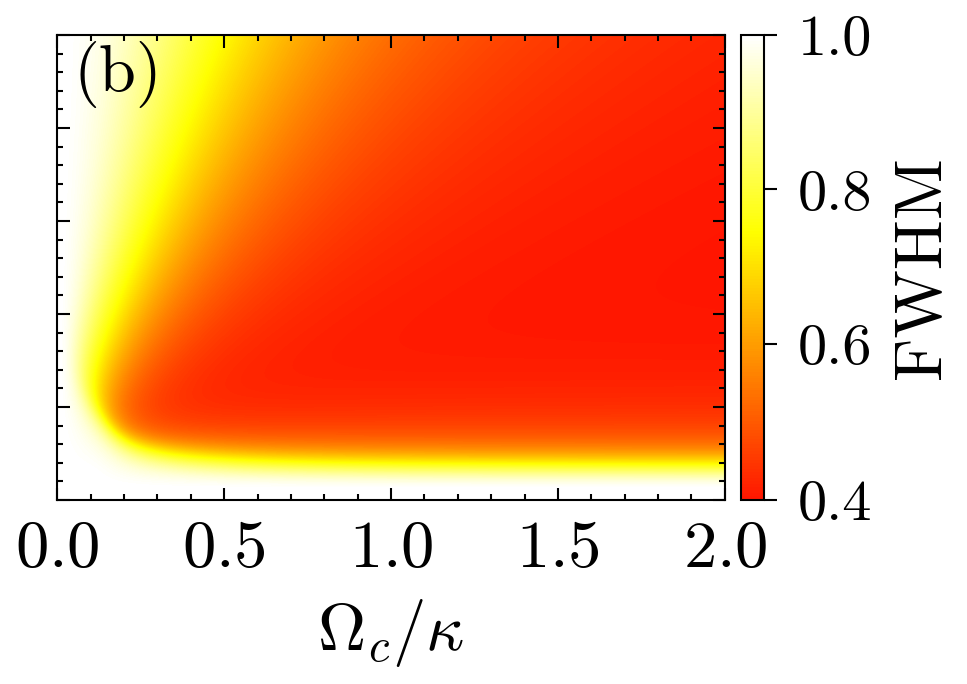}
\caption{FWHM as a function of $g/\kappa$ and $\Omega_c/\kappa$ for $N_{\text{at}}=1$. Here we set the probe field strength as (a) $\varepsilon=\sqrt{0.01}\kappa$ and (b) $\varepsilon=\sqrt{0.1}\kappa$, with spontaneous atomic decay rates $\Gamma_{31}=\Gamma_{32}=0.5\kappa$.}
\label{fig:map} 
\end{figure}

The atom--field coupling strength $g$ sets the cooperativity and determines how efficiently a single emitter can modify the cavity transmission. In the context of our analytical model (Equation~\ref{eq:fwhm_polynomial}), the coupling $g$ directly drives the collective coupling parameter $A \propto g^2$, dictating the strength of the dispersive response that narrows the resonance. A global view of the linewidth dependence on $(g,\Omega_c)$ is shown in Figure~\ref{fig:map} for two probe strengths. Comparing Figure~\ref{fig:map}a and Figure~\ref{fig:map}b, increasing the probe strength drastically reduces the region in parameter space where strong narrowing is observed. The empty-cavity response is recovered for smaller $g$ or for the limiting values of $\Omega_c$, while significant linewidth narrowing requires sufficiently strong coupling and an intermediate control field.

\subsubsection{Low-excitation regime}
For sufficiently weak probe fields, a single atom is capable of drastically modifying the transmitted field, allowing one to reach a very narrow FWHM. This behavior is visible in Figure~\ref{fig:map}a, where we plot the FWHM as a function of $g/\kappa$ and $\Omega_c/\kappa$ for a weak driving $\varepsilon = \sqrt{0.01}\kappa$. As the probe strength $\varepsilon$ is reduced toward the strict linear regime, the maximum achievable narrowing approaches the fundamental non-zero lower bound $F_{\text{min}}$ derived in Appendix~\ref{app:analytical_fwhm}.

This connection between the linewidth behavior and the underlying atom-field dynamics can be further clarified by analyzing the atomic populations as a function of $g$, as shown in Figure~\ref{fig:projection_FWHM}. In these simulations, the probe detuning is set to the edge of the transparency window, $\Delta_p = \text{FWHM}/2$, for each set of parameters. This choice probes the system at a fixed transmission level (half-maximum), where absorption is significant. At a moderate excitation, \textit{e.g.}, with $\varepsilon = \sqrt{0.1}\kappa$ (which already allows a maximum average photon number of $0.1$), Figure~\ref{fig:projection_FWHM}a shows how the transparency window narrows as the atom--field interaction strength $g$ increases. This enhancement in the coupling strength promotes photon absorption in the cavity, leading to a reduction in its linewidth, and a corresponding increase in the atomic population $\langle \sigma_{22} \rangle$ in the ground state $\ket{2}$. Notably, the population in the excited state $\langle \sigma_{33} \rangle$ remains negligible, confirming the effectiveness of EIT in suppressing atomic excitation even when small multiphoton components begin to appear. This nonlinear linewidth behavior has been experimentally observed in similar systems \cite{Wang:00}.

\subsubsection{High-excitation regime}
\begin{figure}[!htb]
\centering
   \includegraphics[width = 1.0\textwidth, clip]{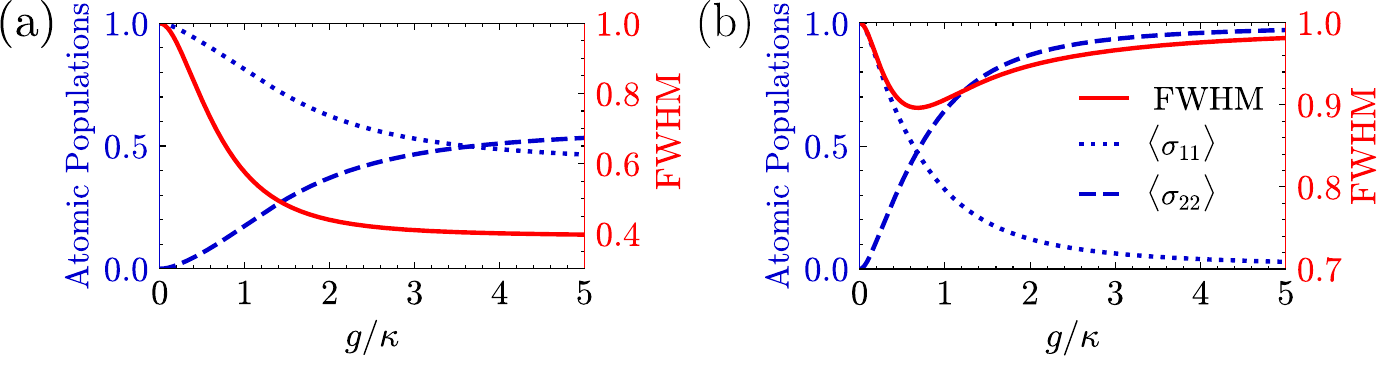}
    \caption{Atomic population $\langle \sigma_{11} \rangle$ and $\langle \sigma_{22} \rangle$ of states $\ket{1}$ and $\ket{2}$, and FWHM as a function of the normalized atom-field coupling strength $g/\kappa$ for $N_{\mathrm{at}}=1$, $\Omega_c=\kappa$ and $\Gamma_{31}=\Gamma_{32}=0.5\kappa$. The probe detuning is set to $\Delta_p=\text{FWHM}/2$ for each point. Two distinct regimes are shown: (a) moderate excitation ($\varepsilon=\sqrt{0.1}\kappa$) and (b) high excitation ($\varepsilon=\sqrt{1.0}\kappa$). The excited state population $\langle \sigma_{33} \rangle$ is negligible in both cases and is therefore not shown.}
    \label{fig:projection_FWHM}
\end{figure}

In contrast, the high-excitation regime ($\varepsilon=\sqrt{1.0}\kappa$), shown in Figure~\ref{fig:projection_FWHM}b, reveals a entirely different physical picture. Here, as $g$ increases, the enhanced cooperativity facilitates a highly efficient, irreversible population transfer from state $\ket{1}$ to state $\ket{2}$ due to incoherent Raman pumping. For large $g$, the atom is almost entirely trapped in the state $\ket{2}$. In our analytical framework, this corresponds to $P_1 \to 0$ ($A \to 0$), effectively decoupling the atom from the cavity mode since the EIT transition is empty. Consequently, the linewidth broadens and fully approaches the empty-cavity limit ($F \to \kappa$). This breakdown can be understood in terms of the critical photon number. For $g=5.0\kappa$, $n_c \approx 0.02$, while the average photon number in the cavity is close to $1.0$. The intracavity field is thus overwhelmingly strong enough to saturate the atom and drive it into a dark state that no longer interacts with the probe, leading to a complete loss of control over the linewidth.

\section{Quantum limit of linewidth narrowing in the few-atom regime}\label{sec:few_atom_regime}

Having established the single-atom behavior ($N_{\text{at}}=1$), we now turn to the few-atom regime and study how an ensemble of $N_{\text{at}}>1$ identical, noninteracting atoms modifies the cavity-EIT response. Since the Hamiltonian and master equation in Section~\ref{sec:PHYSICAL SYSTEM AND MODEL} are formulated for an arbitrary $N_{\text{at}}$, in this section we simply vary the atom number and analyze the resulting changes in photon statistics and linewidth control. For a uniformly coupled ensemble, the relevant symmetric collective excitation couples to the cavity mode with an effective strength $g_{N_{\text{at}}}=g\sqrt{N_{\text{at}}}$.

\subsection{Photon statistics at the optimal operating point}

\begin{figure}[!htb]
\begin{centering}
\includegraphics[width = 1.0\textwidth, clip]{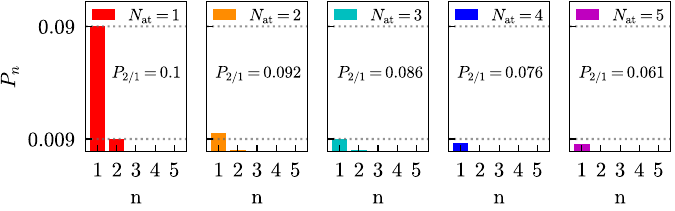}
\par\end{centering}
\caption{Intracavity photon number distribution $P_n$ for $N_{\text{at}}=1$ to $5$ at the operating point that minimizes the linewidth, together with the ratio $P_{2/1}=P_2/P_1$ as an indicator of multiphoton suppression. Here we set $\varepsilon=\sqrt{0.1}\kappa$, $g=5\kappa/\sqrt{N_{\text{at}}}$, and $\Gamma_{31}=\Gamma_{32}=0.5\kappa$. For each $N_{\text{at}}$, $\Omega_c$ and $\Delta_p$ are tuned to achieve $\text{FWHM}_{\text{min}}$, with $\Delta_p=\text{FWHM}_{\text{min}}/2$.}
\label{fig:statistics} 
\end{figure}

We first analyze how the intracavity photon statistics evolve with $N_{\text{at}}$ at the operating point that minimizes the linewidth. Figure~\ref{fig:statistics} shows the photon number distribution $P_n$ for $N_{\text{at}}=1$ to $5$, obtained in a moderate-excitation regime ($\varepsilon=\sqrt{0.1}\kappa$). To compare different atom numbers, we choose $g=5\kappa/\sqrt{N_{\text{at}}}$, so that $g_{N_{\text{at}}}=g\sqrt{N_{\text{at}}}=5\kappa$ remains constant. For each $N_{\text{at}}$, the control field $\Omega_c$ and the probe detuning $\Delta_p$ are tuned to the minimum-linewidth point, with $\Delta_p=\text{FWHM}_{\text{min}}/2$.

Even for a single atom, the field is already strongly suppressed and increasing $N_{\text{at}}$ further reduces the weight of all photon components. A simple indicator of multiphoton suppression is the ratio $P_{2/1}=P_2/P_1$ (annotated in each panel of Figure~\ref{fig:statistics}), which decreases monotonically with $N_{\text{at}}$. This trend signals an increasingly nonlinear, non-classical intracavity field: multiphoton events are progressively inhibited as more atoms participate in the collective absorption. Physically, this is the regime in which the atom-cavity system is nonlinear at the single-(or few-)photon level: high cooperativity ($C\gg 1$) implies that much less than one atom is sufficient to appreciably modify the cavity response, while a small critical photon number ($n_c\leq 1$) implies that a single photon already probes the nonlinearity \cite{RevModPhys.Rempe}. In contrast, in a weak-coupling situation ($g<\kappa,\Gamma$), the photon statistics would remain close to the empty-cavity distribution and would be only weakly affected by adding atoms.

\subsection{Stepwise narrowing and semiclassical benchmark}

\begin{figure}[!htb]
\includegraphics[width = 1.0\textwidth, clip]{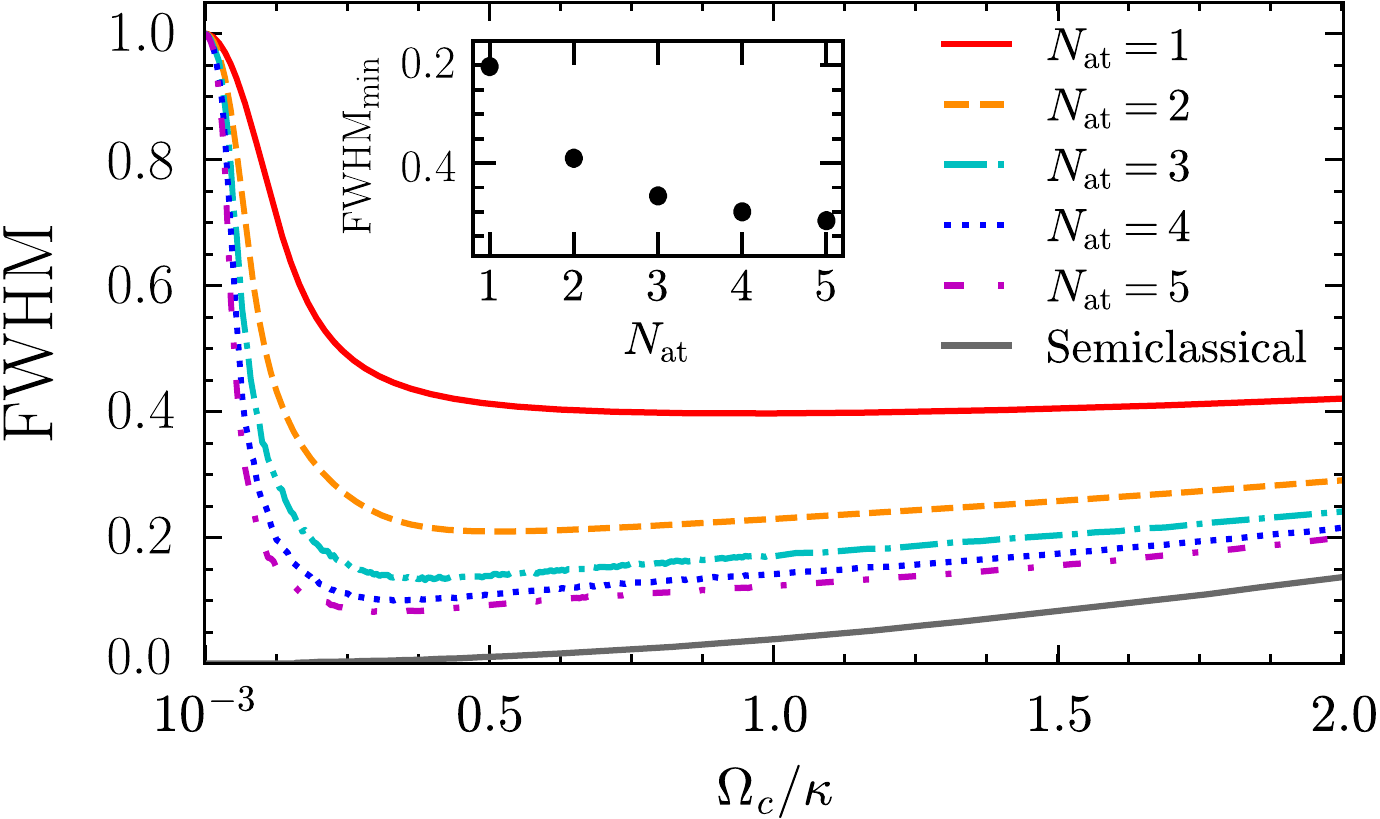}
\caption{FWHM of the cavity transmission as a function of normalized Rabi frequency of control field $\Omega_c/\kappa$. We consider $N_{\text{at}}=1$ to $5$ for the quantum model and $N_{\text{at}}=1000$ using the semiclassical approximation. The inset shows the minimum FWHM values for each number of atoms confined in the cavity. Here we set $\varepsilon=\sqrt{0.1}\kappa$, $g=5\kappa/\sqrt{N_{\text{at}}}$ and $\Gamma_{31}=\Gamma_{32}=0.5\kappa$.}
\label{fig:EIT_FWHM} 
\end{figure}

This comparison is particularly important because previous linewidth narrowing studies establish the expected collective trend in the strictly linear-response EIT regime, whereas here we ask whether this narrowing remains unbounded once the field is treated quantum mechanically and the probe amplitude is finite. We now connect the intracavity statistics directly to the linewidth control. Figure~\ref{fig:EIT_FWHM} shows the FWHM as a function of $\Omega_c/\kappa$ for $N_{\text{at}}=1$ to $5$. In the regime considered here, the single-atom case is particularly sensitive to saturation: the non-negligible two-photon component limits how efficiently one atom can enforce the EIT-induced narrowing, resulting in a comparatively broad transparency feature across $\Omega_c$. 

A striking feature in Figure~\ref{fig:EIT_FWHM} is the stepwise (``staircase'') reduction of the minimum achievable linewidth with $N_{\text{at}}$ (inset of Figure~\ref{fig:EIT_FWHM}). This behavior reflects the discrete nature of the coupled atom--cavity excitation structure. In the ideal linear-response limit discussed in Section~\ref{subsec:control_field_Oc}, the trend is consistent with the familiar $\text{FWHM}\propto \Omega_c^2/(g^2 N_{\text{at}})$ scaling \cite{mucke2010electromagnetically}. However, Figure~\ref{fig:EIT_FWHM} demonstrates that, in the few-photon quantum regime, the narrowing is ultimately bounded and does not evolve smoothly toward an arbitrarily small linewidth for a finite probe. Importantly, the optimal operating point corresponds to the control-field strength that best enforces destructive interference before nonlinear Raman transitions fully trap the population.

To unveil the physical origin of this fundamental bound, we benchmark the few-atom exact quantum results against the many-atom trend computed via the semiclassical approximation~\cite{PhysRevA.2.336} (Appendix \ref{SC}) for $N_{\text{at}}=1000$. The semiclassical curve in Figure~\ref{fig:EIT_FWHM} displays an almost vanishing linewidth as $\Omega_c \to 0$, suggesting that the narrowing has no lower bound. However, this ``unbounded'' behavior is a direct consequence of the mean-field factorization: replacing the cavity operators by a coherent amplitude inherently ignores the discrete photon nature of the field. 

In the full quantum model, driving the cavity with a finite coherent probe unavoidably populates higher-order multiphoton states ($n \ge 2$). As demonstrated analytically by our dressed-state evaluation in Appendix~\ref{app:analytical_fwhm}, these multiphoton components are highly dissipative. The decay rate of the dark states belonging to higher excitation manifolds causes them to remain lossy even when $\Omega_c \to 0$. The presence of these discrete multiphoton excitations directly broadens the EIT transmission window and destroys the idealized single-excitation dark resonance. This establishes the fundamental physical mechanism behind the quantum limit.

As $N_{\text{at}}$ increases, the enhanced collective cooperativity strengthens the photon blockade mechanism, which strongly suppresses these detrimental multiphoton states (as evidenced by the rapid drop in $P_{2/1}$ in Figure~\ref{fig:statistics}). By reducing the statistical weight of the $n \ge 2$ broadening channels, the quantum limit to the FWHM is progressively lowered, giving rise to the stepwise narrowing trend. In the macroscopic limit ($N_{\text{at}} \gg 1$), the multiphoton probability is heavily inhibited, and the system asymptotically suppresses the quantum limits, recovering the semiclassical linear-response prediction.

\section{Conclusions}\label{sec:CONCLUSIONS}

We have presented a quantum theoretical analysis of linewidth control in cavity EIT, with emphasis on the transition from a single emitter to the few-atom regime. Using the Lindblad master equation framework, we mapped how the FWHM of the central transparency resonance depends on the control-field strength $\Omega_c$, probe amplitude $\varepsilon$, coupling strength $g$, and atom number $N_{\text{at}}$. For a single atom, we found that linewidth control is fundamentally constrained by the degree of intracavity excitation. In the strictly low-excitation regime, we derived an exact analytical cubic polynomial that perfectly captures the EIT-induced narrowing and its physical lower bounds. However, for higher probe powers, saturation limits how efficiently one atom can reshape the transmitted field. The global $(g,\Omega_c)$ maps further highlight that strong narrowing requires sufficiently large cooperativity and an intermediate control field. Outside this optimal operating window, incoherent Raman pumping decouples the atom, and the response safely reverts to the empty-cavity linewidth.

In the few-atom regime, we showed that increasing $N_{\text{at}}$ enables additional narrowing and leads to a distinct stepwise reduction of the minimum achievable linewidth. Crucially, our main result is that, for any finite probe power, linewidth narrowing is ultimately bounded by a fundamental quantum limit. By contrasting the full quantum model with a semiclassical approximation, we established an operational link between this fundamental bound and the coherent excitation of higher-order multiphoton states. As demonstrated analytically, these multiphoton components possess intrinsically larger decay rates that destroy the ideal single-excitation EIT dark state. Increasing $N_{\text{at}}$ enhances the collective cooperativity, which acts as a strong photon blockade mechanism. This effect progressively suppresses the detrimental multiphoton excitations and lowers the quantum barrier, driving the system step-by-step toward the idealized mean-field macroscopic behavior.

These results clarify the quantum nature of controllability in cavity EIT and provide concrete, exact analytical boundaries for device optimization. In practice, engineering ultra-narrow transmission windows requires operating near the optimal $\Omega_c$ set by the interplay of interference and excitation, and critically balancing the continuous probe power against the onset of multiphoton dissipative channels. Understanding this parameter dependence is directly relevant to the design of cavity-based narrowband optical filters and highly coherent cavity-assisted light-matter interfaces for quantum technologies, including quantum networking and quantum-memory applications \cite{specht2011single, LEI_Review_EIT}.

\medskip
\textbf{Acknowledgements} \par 
This study was financed in part by the Coordenação de Aperfeiçoamento de Pessoal de Nível Superior - Brasil (CAPES) - Finance Code 001, the São Paulo Research Foundation (FAPESP), Brasil (Process Numbers \#2019/11999-5, \#2022/00209-6, \#2022/05551-4 and \#2024/02604-5), the Brazilian National Council for Scientific and Technological Development (CNPq, Grant No.~311612/2021-0, No.~141247/2018-5, and No.~405712/2023-5), and CAPES-STINT BR2018-8054.

% Appendix
\medskip
\appendix

\section{The semiclassical approximation}
\label{SC}
 
Simulation of quantum many body systems is a great challenge due to the Hilbert space dimension being too large when increasing $N_{\text{at}}$. One way of avoiding such hardship is to take the semiclassical approximation, which can be applied in the weak atom-field coupling regime ($g<\kappa, \Gamma)$. In Figure~\ref{fig:EIT_FWHM} the simulation for $N_{\text{at}}=1000$ showed an almost vanishing linewidth for $\Omega_c<\kappa$, however the coupling strength employed does not satisfy the condition of this approximation, therefore it does not compromise the analysis.

The time evolution of the time-dependent operators can be obtained via $\langle \dot{O}\rangle = \text{Tr}\left({\dot{\rho} O} \right)$. By approximating the atom-field correlations as $\left\langle a S_{ij} \right\rangle \approx \alpha \left\langle S_{ij} \right\rangle$, where $\alpha$ represents the time-dependent amplitude of the cavity field, we effectively simplify the system by neglecting all higher order (quantum) correlations of the field. By applying this approximation in the set of nonlinear coupled differential equations as previously discussed in \cite{PhysRevA.2.336}, we obtain
    \begin{align}
    \dot{\alpha} &= i\left[\left(\Delta_p+i\frac{\kappa}{2}\right)\alpha - g\left \langle  S_{13} \right \rangle - \varepsilon\right] ,\\%<a>
    \langle \dot{S}_{12} \rangle &= i\Delta_p \langle S_{12} \rangle -i\Omega_c \langle S_{13} \rangle + ig \alpha \langle S_{32} \rangle, \\ %<S_12>
     \langle \dot{S}_{13} \rangle &= i\left[\Delta_p+ \frac{i}{2}\left(\Gamma_{31} + \Gamma_{32}\right)\right] \langle S_{13} \rangle - i\Omega_c \langle S_{12}  \rangle +i g \alpha\left(\langle S_{33} \rangle - \langle S_{11} \rangle \right)  , \\ %<S_13>
      \langle \dot{S}_{23} \rangle &= - \dfrac{1}{2}(\Gamma_{31} + \Gamma_{32}) \langle S_{23} \rangle -i g\alpha\langle S_{21} \rangle +i \Omega_c\left( \langle S_{33} \rangle -\langle S_{22} \rangle \right) , \\%<S_23>
      \langle \dot{S}_{11} \rangle &= -ig\alpha^*\left \langle S_{13} \right \rangle + i g \alpha \left \langle S_{31} \right \rangle + \Gamma_{31}\left \langle S_{33} \right \rangle \, , \\%<S_11>
      \langle \dot{S}_{22} \rangle &= -i\Omega_c\left \langle S_{23} \right \rangle + i \Omega_c \left \langle S_{32} \right \rangle + \Gamma_{32}\left \langle S_{33} \right \rangle  ,\\%<S_22> 
       \langle \dot{S}_{33}  \rangle &= -  \langle \dot{S}_{11}  \rangle -   \langle \dot{S}_{22}  \rangle ,%<S_33>
\end{align}
where it is possible to numerically solve them (and their respective hermitian conjugates) for long enough times in order to find the system's steady state.

\section{Analytical linewidth solution} 
\label{app:analytical_fwhm}

To gain a deeper physical understanding of the cavity-EIT linewidth narrowing and its ultimate limits, we derive an analytical expression for the FWHM using a semiclassical mean-field approach combined with the adiabatic elimination of the excited state.

\subsection{Semiclassical field amplitude} 

Building upon the semiclassical approach and the set of differential equations derived in Appendix~\ref{SC}, we apply the steady-state condition ($\langle\dot{O}\rangle = 0$). From the equation of motion for the cavity field amplitude $\alpha$, we can readily isolate it as:
\begin{equation} 
    \alpha(\Delta_p) = \frac{\varepsilon}{\Delta_p + i\frac{\kappa}{2} + \chi_{\text{at}}(\Delta_p)}, 
\end{equation}
where $\chi_{\text{at}}(\Delta_p)= -g\langle S_{13}\rangle / \alpha$ is the effective atomic susceptibility. To find this susceptibility, we evaluate the steady-state atomic coherences from Appendix~\ref{SC}. Assuming the probe field is sufficiently weak such that the excited state population and the coherence $\langle S_{32}\rangle$ are negligible ($\langle S_{33}\rangle \approx 0$, $\langle S_{32}\rangle \approx 0$), the equation for $\langle S_{12} \rangle$ yields:
\begin{equation}
    \langle S_{12} \rangle = \frac{\Omega_c}{\Delta_p} \langle S_{13} \rangle.
\end{equation}
Substituting this relation into the steady-state equation for $\langle S_{13} \rangle$, we obtain:
\begin{equation}
    0 = i\left(\Delta_p + i\frac{\Gamma}{2}\right) \langle S_{13} \rangle - i\frac{\Omega_c^2}{\Delta_p} \langle S_{13} \rangle - ig\alpha \langle S_{11} \rangle,
\end{equation}
where $\Gamma = \Gamma_{31} + \Gamma_{32}$. Solving for $\langle S_{13} \rangle$ and defining $P_1 = \langle S_{11} \rangle$ as the steady-state probability of finding the atom in the ground state $\ket{1}$, we find:
\begin{equation}
    \langle S_{13} \rangle = \frac{g \alpha P_1 \Delta_p}{\Delta_p^2 + i\Delta_p\frac{\Gamma}{2} - \Omega_c^2}.
\end{equation}
This directly leads to the effective atomic susceptibility:
\begin{equation}
    \chi_{\text{at}}(\Delta_p) = \frac{g^2P_1 \Delta_p}{\Omega_c^2 - \Delta_p^2 - i\Delta_p\frac{\Gamma}{2}}.
\end{equation} 

\subsection{Ground-state population} 

To determine $P_1$, we evaluate the populations within the ideal EIT dark state. Applying a displacement transformation $a \to a + \alpha$ to the system Hamiltonian and truncating the dynamics to the $n$-excitation subspace with the basis $\{\ket{n,1}, \ket{n-1,3}, \ket{n-1,2}\}$, the effective non-Hermitian Hamiltonian takes the following matrix form:
\begin{equation}
    H_{\text{eff}}^{(n)} = \begin{pmatrix} 0 & g\sqrt{n} & 0 \\ g\sqrt{n} & 0 & \Omega_c \\ 0 & \Omega_c & 0 \end{pmatrix}.
\end{equation}
Diagonalizing this matrix reveals a zero-energy dark state:
\begin{equation}
    \ket{D_n} = \frac{\Omega_c \ket{n,1} - g\sqrt{n} \ket{n-1,2}}{\sqrt{\Omega_c^2 + g^2 n}},
\end{equation}
along with two lossy bright states $\ket{B_\pm}$. In the optimal EIT transmission window, the system is optically pumped into this dark state. Thus, the probability $P_1$ of finding the atom in the ground state $\ket{1}$ is simply given by the projection of the dark state onto $\ket{n,1}$:
\begin{equation} 
    P_1 = |\braket{n,1}{D_n}|^2 = \frac{\Omega_c^2}{\Omega_c^2 + g^2 n}, 
\end{equation} 
where $n=|\alpha|^2$ is the intracavity photon number. Remarkably, calculating the effective cavity decay rate for this dressed state via Fermi's Golden Rule using the jump operator $C = \sqrt{\kappa}a$ yields $\gamma_D = \kappa \bra{D_1} a^\dagger a \ket{D_1} = \kappa \Omega_c^2/(\Omega_c^2 + g^2)$ for $n=1$. This decay rate results in the exact same spectral width scaling obtained by simply evaluating the population projection $P_1$. To determine the analytical FWHM, we evaluate this saturation at the half-maximum condition of the empty cavity, $n_\text{FWHM}=2(\varepsilon/\kappa)^2$.

\subsection{Exact cubic polynomial for the FWHM} 

The normalized cavity transmission is proportional to $|\alpha(\Delta_p)|^2$. In the central transparency window, the control field induces destructive interference, making the imaginary part of the susceptibility (absorption) negligible, $\text{Im}(\chi_{\text{at}})\approx0$. The half-maximum condition of the transmission peak occurs when the real dispersive term doubles the squared denominator: 
\begin{equation} 
    \left( \Delta_p + \frac{g^2  P_1 \Delta_p}{\Omega_c^2 - \Delta_p^2} \right)^2 = \left( \frac{\kappa}{2} \right)^2. 
\end{equation} 
Taking the positive root and substituting the detuning by the half-width $\Delta_p=F/2$, where $F$ is the FWHM, we obtain: 
\begin{equation}
    \frac{F}{2} + \frac{g^2  P_1 (F/2)}{\Omega_c^2 - (F/2)^2} = \frac{\kappa}{2}. 
\end{equation} 
By defining the effective saturated coupling parameter $A=g^2P_1=g^2\Omega_c^2/(\Omega_c^2+g^2n_\text{FWHM})$ and rearranging the terms by orders of $F$, we arrive at the exact cubic polynomial for the FWHM: 
\begin{equation} 
    P(F, \Omega_c) = F^3 - \kappa F^2 - 4(\Omega_c^2 + A)F + 4\kappa\Omega_c^2 = 0. \label{eq:app_cubic_polynomial} 
\end{equation}
The linewidth of the central EIT resonance corresponds to the smallest real positive root of this polynomial. In the limit $\Omega_c\rightarrow 0$, the population is entirely pumped to $\ket{2}$ ($A\rightarrow0$), and the polynomial trivially reduces to $F^2(F-\kappa)=0$, recovering the empty-cavity linewidth $F=\kappa$.

\subsection{Minimum FWHM and Validity} 

As the FWHM converges to $\kappa$ for both very strong ($\Omega_c \gg g$) and very weak ($\Omega_c\rightarrow 0$) control fields, there must exist a global minimum. The minimum achievable linewidth $F_\text{min}$ and its corresponding optimal control field $\Omega_{c,\text{opt}}$ can be found analytically using implicit differentiation on Eq.~\eqref{eq:app_cubic_polynomial}. Setting $dF/d\Omega_c=0$, we get: 
\begin{equation} 
    -8\Omega_c F_{\text{min}} - 4 \frac{dA}{d\Omega_c}F_{\text{min}} + 8\kappa\Omega_c = 0, 
\end{equation} 
which isolates the minimum FWHM as: 
\begin{equation}
    F_{\text{min}} = \frac{2\kappa\Omega_c}{2\Omega_c + \frac{dA}{d\Omega_c}}.
\end{equation} 
Calculating the derivative $dA/d\Omega_c$ and simplifying, we find the fundamental lower bound: 
\begin{equation}
    F_{\text{min}} = \frac{\kappa}{1 + \frac{g^4 N_{\text{at}} n_{\text{FWHM}}}{(\Omega_{c,\text{opt}}^2 + g^2 n_{\text{FWHM}})^2}}. 
\end{equation} 

This closed form expression formally proves that, for any finite probe strength ($\varepsilon >0 \Rightarrow n_{\text{FWHM}}>0$), the denominator is strictly greater than 1. Consequently, the minimum FWHM is fundamentally bounded ($F_{\text{min}} <\kappa$) and cannot reach zero. It is crucial to emphasize that this approximation and the resulting Eq.~\eqref{eq:app_cubic_polynomial} are strictly valid only in the low-excitation limit, specifically for very weak probe fields $\varepsilon \lesssim  \sqrt{0.001}\kappa$. For higher probe strengths, multiphoton components ($n \ge 2$) become significantly populated. As derived from Fermi's Golden Rule, the decay rate for these higher-order dressed states scales as $\gamma_{D,n} \to \kappa(n-1)$ when $\Omega_c \to 0$, which inevitably broadens the linewidth and makes it diverge from the $n=1$ analytical prediction. Finally, it should be noted that this approach yields a semi-analytical solution, since extracting the FWHM ultimately relies on numerically finding the roots of a cubic polynomial.

% References
\medskip

\bibliographystyle{MSP}
\bibliography{bib}

\end{document}